\documentclass[aps,pra,showpacs,amsmath,twocolumn]{revtex4}
\usepackage[dvips]{graphicx}

\begin{document}

\title{Dynamics of saturated Bragg diffraction in a stored light grating in cold atoms}

\author{D. Moretti$^{(1)}$, N. Gonzalez$^{(2)}$, D. Felinto$^{(1)}$, and J. W. R. Tabosa$^{(1)}$}
\affiliation{$^{(1)}$ Departamento de F\'isica, Universidade
Federal de Pernambuco, Cidade Universit\'aria, 50670-901 Recife,
PE, Brazil}

\affiliation{$^{(2)}$Institut de Ciencies Fotonique (ICFO),
Barcelona, Spain}

\date{\today}

\begin{abstract}
We report on a detailed investigation of the dynamics and the
saturation of a light grating stored in a sample of cold cesium
atoms. We employ Bragg diffraction to retrieve the stored optical
information impressed into the atomic coherence by the incident
light fields. The diffracted efficiency is studied as a function
of the intensities of both writing and reading laser beams. A
theoretical model is developed to predict the temporal pulse shape
of the retrieved signal and compares reasonably well with the
observed results.
\end{abstract}


\maketitle

\section{Introduction}

The storage of light information in an atomic ensemble is a well
understood phenomenon which has a promising prospect for
application both in classical and quantum information processing
\cite{Lukin03}. The light storage (LS) phenomenon allows us to
obtain later information on a previously stored light pulse, as
well as to manipulate the stored information. As it was first
proposed, LS in an electromagnetically induced transparency (EIT)
medium \cite{Harris97, Fleischhauer05} is described in terms of a
mixed two component light-matter excitation, called dark state
polariton (DSP), where each component of the excitation can be
externally controlled \cite{Freischhauer00}. To date, several
experimental observations of these effects were realized in
different systems \cite{Phillips01, Liu01, Zibrov02, Mair02,
Wang05}.

Alternatively, the LS process can also be described as being due
to the creation of a spatially dependent ground states coherence
that contains respectively the information on the amplitude and
phase of a light pulse and which survives after the switching-off
of the incident light. Using this simpler picture, we have
recently demonstrated the storage of a polarization light grating
into an atomic coherence via a backward four-wave mixing
configuration \cite{Tabosa07}. Others schemes have also been
recently employed to store spatial structures (images) in atomic
vapors \cite{Shuker07, Pugatch07}. For instance, a light vortex
was stored in a hot vapor for hundreds of microseconds
\cite{Pugatch07}.

In this work we present experimental and theoretical investigation
on the dynamics of light grating stored in an EIT medium
associated with a degenerate two-level system. The dependence of
the stored light grating with the intensities of the incident
writing and reading beams is investigated. Bragg diffraction into
the stored grating is employed to probe its dynamics under
different experimental conditions. The demonstration of the
reversible storage and the manipulation of the spatial light phase
structure stored into the atomic ensemble, and its extension to
include beams carrying orbital angular momentum, would be of great
importance to demonstrate the manipulation of quantum information
encoded in a higher dimensional state space \cite{Padgett04,
Barreiro03}. Moreover, the storage of this light grating opens up
the possibility to investigate the generation of correlated
photons pairs in a previously coherently prepared atomic ensemble
\cite{Balic05}.

\vspace{-0.3 cm}
\section{Theoretical model}

\begin{figure}[htb]
\includegraphics[angle=0,scale=0.45]{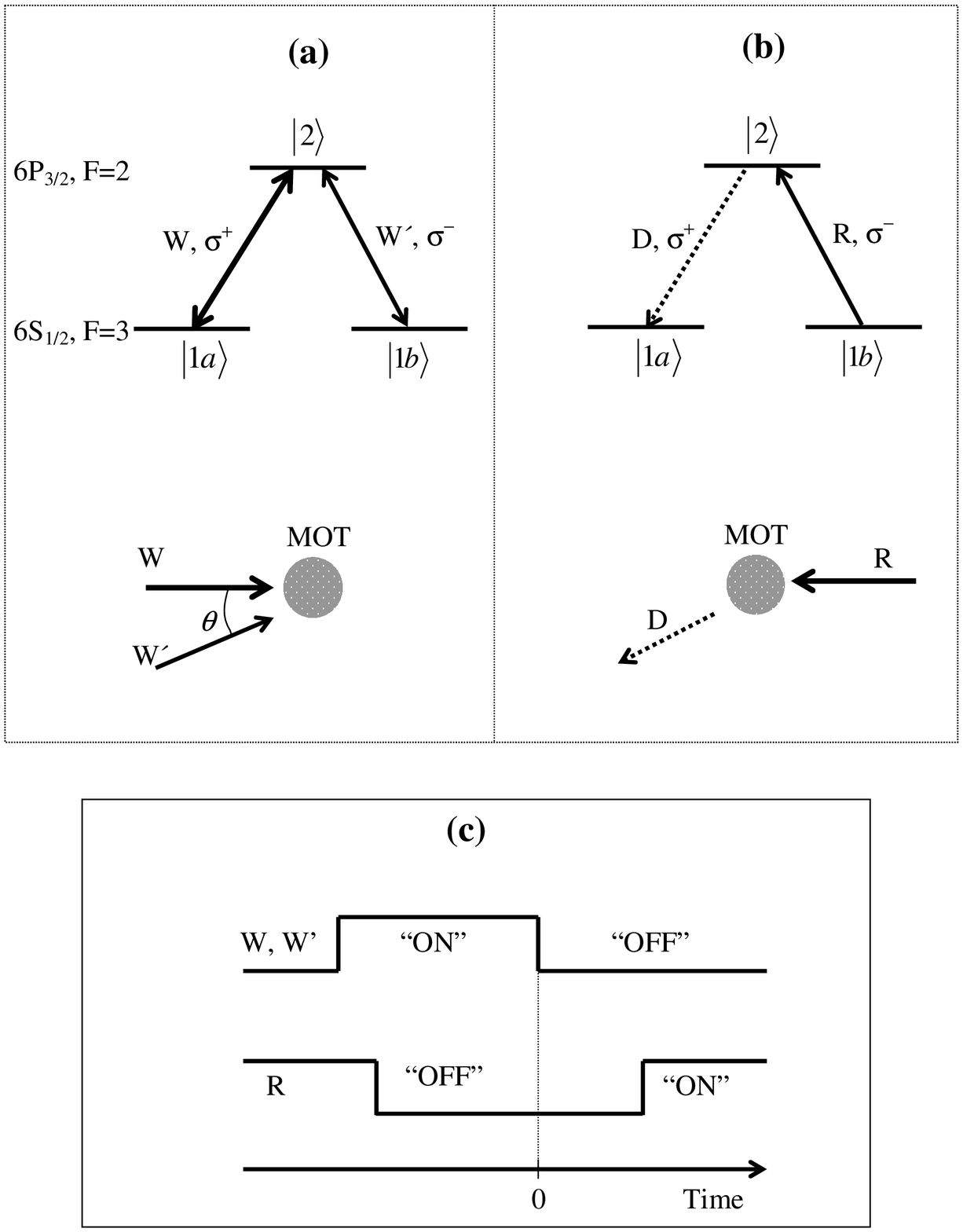}
\vspace{0.1 cm} \caption{(a) Simplified Zeeman level scheme,
showing the coupling and the propagation directions of the grating
writing beams ($W$ and $W^{\prime}$) and (b) the coupling and the
propagation direction of the reading ($R$) and diffracted ($D$)
beams. The beams $W$ and $W^{\prime}$ make a small angle $\theta$
and are circularly polarized with opposite handedness, while the
beam $R$ is counterpropagating to the beam $W$ and have a circular
polarization opposite to this beam. The diffracted beam is
detected in a direction opposite to the beam $W^{\prime}$.(c) The
switching time sequence for the writing and reading beams.}
\label{setup}
\end{figure}

We consider an ensemble of cold atoms excited by three different
fields: two writing ($W$ and $W^{\prime}$) and  one reading ($R$)
laser pulses. The atomic ensemble can be well approximated by a
set of degenerate two-level atoms, with a ground-state manifold
composed of two degenerate states ($|1a\rangle$ and $|1b\rangle$)
and the excited-state manifold having a single state
($|2\rangle$). As illustrated in Fig.~\ref{setup}, the
ground-state degeneracy corresponds to the Zeemam degeneracy of
atomic cesium in the experiment. In this way, the different levels
are connected by fields of different polarizations with respect to
the atom. We consider fields $W^{\prime}$ and $R$ having
$\hat{\sigma}^-$ polarization and field $W$ having
$\hat{\sigma}^+$ polarization. $W^{\prime}$ and $R$ excite then
the transition $1b \rightarrow 2$, and $W$ the transition $1a
\rightarrow 2$.

The fields $W$ and $W^{\prime}$ propagate in different directions,
corresponding to a small angle $\theta$ between them. The $R$
field is counter-propagating with respect to $W$. The signal we
want to model corresponds to the diffraction of the $R$ field in
the spatial grating formed by fields $W$ and $W^{\prime}$. In the
case of cw excitation of the ensemble, this signal corresponds to
the well-know conjugated signal in four-wave mixing (FWM)
processes \cite{Lezama01}. Here we call it $D$ field (see
Fig.~\ref{setup}b).

We use this FWM configuration to store and later retrieve a
coherence grating written in the atomic ensemble. In order to
address this coherence storage process, we use a specific time
sequence for the pulsed excitation of the ensemble, see
Fig.~\ref{setup}c. First we prepare the sample by exciting it with
the two, long writing pulses. In this writing process, the goal is
to leave the system in its stationary state. Then we turn off the
writing beams, and wait a certain amount of time, the storage
time, before turning the reading pulse on. This reading pulse
stays on also for a long time, enough to extract the whole stored
grating from the ensemble. A field-$D$ pulse is then generated
during the read process. In the following theoretical analysis, we
want to model and study this field-$D$ generation process in
detail, considering the three-level-atom approximation discussed
above.

\subsection{Grating formation and storage}
\label{write}

Consider an atom excited by two writing beams: one, $\vec{E}_W$,
propagating in the $z$ direction and the other,
$\vec{E}_{W^{\prime}}$, forming an angle $\theta$ with
$\vec{E}_W$. The fields $\vec{E}_W$ and $\vec{E}_{W^{\prime}}$
have orthogonal circular polarizations $\hat{\sigma}^+$ and
$\hat{\sigma}^-$, respectively. We consider small enough angles so
that we can assume, to a good approximation, the polarization of
$\vec{E}_{W^{\prime}}$ as being $\hat{\sigma}^-$ on the same state
basis in which $\vec{E}_W$ is $\hat{\sigma}^+$. We can then write
\begin{subequations}
\begin{align}
\vec{E}_W &= {\cal E}_W (\vec{r}) e^{i(k_W z - \omega_W t)} \hat{\sigma}^{+} \;, \\
\vec{E}_{W^{\prime}} &= {\cal E}_{W^{\prime}} (\vec{r})
e^{i(\vec{k}_{W^{\prime}}\cdot\vec{r} - \omega_{W^{\prime}} t)}
\hat{\sigma}^{-} \;,
\end{align}
\label{campo}
\end{subequations}
where ${\cal E}_W (\vec{r})$ and ${\cal E}_{W^{\prime}} (\vec{r})$
represent the transversal modes of each field, and we assumed both
of them having constant intensities. The frequencies of the fields
are $\omega_W$ and $\omega_{W^{\prime}}$, and their wavevectors
are $k_W \hat{z}$ and $\vec{k}_{W^{\prime}}$, respectively. The
energy difference between fundamental and excited levels is $\hbar
\omega_e$.

The system hamiltonian can then be written as
\begin{equation}
\hat{H}(t) = \hat{H}_0 + \hat{V}(t) \;,
\end{equation}
where
\begin{equation}
\hat{H}_0 = \hbar \omega_e |2\rangle \langle 2|
\end{equation}
is the hamiltonian for the free atom and
\begin{align}
\hat{V}(t) =& -\vec{d}_{2,1a}\cdot \vec{E}_W(t)\, |2\rangle
\langle 1a| \nonumber \\ &\;\;\; -\vec{d}_{2,1b}\cdot
\vec{E}_{W^{\prime}}(t)\, |2\rangle \langle 1b| + h.c. \,
\label{interacao}
\end{align}
is the interaction hamiltonian. Defining the Rabi frequencies
\begin{subequations}
\begin{align}
\Omega_W(\vec{r}) &= \frac{id_{2,1a} {\cal E}_W(\vec{r}) e^{ik_W z}}{\hbar}\;, \\
\Omega_{W^{\prime}}(\vec{r}) &= \frac{id_{2,1b}{\cal
E}_{W^{\prime}}(\vec{r}) e^{i\vec{k}_{W^{\prime}}\cdot
\vec{r}}}{\hbar}\;,
\end{align}
\end{subequations}
 and assuming the resonance condition $\omega_W = \omega_{W^{\prime}} = \omega_e$, the whole set of Bloch equations for the system, in the rotating-wave approximation, becomes
\begin{subequations}
\begin{align}
\frac{d\rho_{22}}{dt} &= \left[ \Omega_W \sigma_{1a,2} + \Omega_{W^{\prime}} \sigma_{1b,2} + c.c.\right] - \Gamma_{22} \rho_{22} \;, \\
\frac{d\rho_{1a,1a}}{dt} &= \left[ -\Omega_W \sigma_{1a,2} + c.c.\right] + \Gamma_{1a,1a}^{\prime}\rho_{22} \;, \\
\frac{d\rho_{1b,1b}}{dt} &= \left[ -\Omega_{W^{\prime}}\sigma_{1b,2} + c.c.\right] + \Gamma_{1b,1b}^{\prime}\rho_{22} \;,\\
\frac{d\sigma_{1a,2}}{dt} &= -\Omega_W^* (\rho_{22} - \rho_{1a,1a}) + \Omega_{W^{\prime}}^*\rho_{1a,1b} - \Gamma_{12} \sigma_{1a,2} \;,\\
\frac{d\sigma_{1b,2}}{dt} &= -\Omega_{W^{\prime}}^* (\rho_{22} - \rho_{1b,1b}) + \Omega_W^*\rho_{1b,1a} - \Gamma_{12} \sigma_{1b,2} \;,\\
\frac{d\rho_{1a,1b}}{dt} &= -\Omega_W^* \sigma_{2,1b} -
\Omega_{W^{\prime}} \sigma_{1a,2} - \gamma \rho_{1a,1b} \;,
\end{align}
\end{subequations}
with $\sigma_{1a,2} = \rho_{1a,2} \,e^{-i\omega_W t}$ and
$\sigma_{1b,2} = \rho_{1b,2} \,e^{-i\omega_{W^{\prime}} t}$. The
spontaneous relaxation rates are indicated by $\Gamma_{12}$ and
$\Gamma_{22}$, for the coherence and population decays,
respectively. $\Gamma_{1a,1a}^{\prime}$ and
$\Gamma_{1b,1b}^{\prime}$ indicate the rates at which the
$\rho_{22}$ population decays into the populations $\rho_{1a,1a}$
and $\rho_{1b,1b}$, respectively. For simplicity, in these
equations and in the following, we omit the spatial dependence of
the Rabi frequencies. The ground-state-coherence decay rate
$\gamma$ is introduced to take into account, in an effective way,
the decay induced by residual magnetic fields. Such decay is
usually a result of inhomogeneous broadening in the ensemble of
atoms, each subject to a slightly different magnetic
field~\cite{Felinto_2005}. For the signal we are treating here,
however, this simple model considering the same decay constant for
the whole ensemble is already enough to obtain a good comparison
with the experimental data probing the coherence decay.

After a sufficiently long time, the system reaches a steady
situation in which $d\rho_{kl}/dt = 0$, for all $\rho_{kl}$
density-matrix elements. The steady-state coherence
$\rho_{1a,1b}^e$ between the two ground state levels is then given
by
\begin{equation}
\rho_{1a,1b}^e =
-\frac{\left(\Gamma_{1a,1a}^{\prime}|\Omega_{W^{\prime}}|^2 +
\Gamma_{1b,1b}^{\prime}|\Omega_W|^2 \right)}{A} \;\Omega_W^*
\Omega_{W^{\prime}} \;,
\end{equation}
with
\begin{align}
&A = \nonumber \\ &\;\; \left(\Gamma_{1a,1a}^{\prime}|\Omega_{W^{\prime}}|^2 + \Gamma_{1b,1b}^{\prime}|\Omega_W|^2 \right)\left( \gamma\Gamma_{12} + |\Omega_W|^2 + |\Omega_{W^{\prime}}|^2\right) \nonumber \\
& \;\;\; + 6\gamma |\Omega_{W^{\prime}}|^2|\Omega_W|^2 \;.
\end{align}
We are particularly interested in the situation where $\gamma$ is
very small when compared to any other frequency in the system,
since this corresponds to our experimental condition. In this
limit, note then that the above expression simplifies to
\begin{equation}
\rho_{1a,1b}^e = -\frac{\Omega_W^*
\Omega_{W^{\prime}}}{|\Omega_W|^2 + |\Omega_{W^{\prime}}|^2} \;.
\label{rhoe}
\end{equation}

Once the fields $\vec{E}_W$ and $\vec{E}_{W^{\prime}}$ are turned
off, the coherences in the system evolve according to their
respective decay times. Since $\gamma << \Gamma_{12}$, after a
time $t_s >> 1/\Gamma_{12}$ the stored coherences in the sample
can be well approximated by
\begin{subequations}
\begin{align}
\sigma_{1a,2}^s(t_s) &= 0 \;,\\
\sigma_{1b,2}^s(t_s) &= 0 \;,\\
\rho_{1a,1b}^s(t_s) &= \rho_{1a,1b}^e e^{-\gamma t_s} \;.
\end{align}
\label{rhos}
\end{subequations}

\subsection{Reading}
\label{read}

The stored coherence grating can be extracted from the sample
using a $\hat{\sigma}^-$-polarized third field $\vec{E}_R$
counter-propagating with respect to $\vec{E}_W$:
\begin{equation}
\vec{E}_R = {\cal E}_R (\vec{r}) e^{i(-k_R z - \omega_R t)}
\hat{\sigma}^{-} \;,
\end{equation}
with ${\cal E}_R$, $k_R$, and $\omega_R$ representing the
transversal mode, wavevector, and frequency, respectively, of
field $\vec{E}_R$. After similar considerations as for the
grating-formation process, including the resonance condition
$\omega_e = \omega_R$, and the analogous definition of a third
Rabi frequency
\begin{equation}
\Omega_R(\vec{r}) = \frac{id_{2,1b} {\cal E}_R(\vec{r}) e^{-ik_R
z}}{\hbar}\;,
\end{equation}
the relevant Bloch equations describing the reading process become
\begin{subequations}
\begin{align}
\frac{d\sigma_{1a,2}}{dt} &= \Omega_R^*\rho_{1a,1b} - \Gamma_{12} \sigma_{1a,2} \;,\\
\frac{d\rho_{1a,1b}}{dt} &= - \Omega_R \sigma_{1a,2} - \gamma
\rho_{1a,1b} \;,
\end{align}
\label{reading_01}
\end{subequations}
with $\sigma_{1a,2} = \rho_{1a,2} \,e^{-i\omega_R t}$. Note that
the equations for $\sigma_{1a,2}$ and $\rho_{1a,1b}$ are actually
de-coupled from the rest of the system of Bloch equations.

We are interested in calculating the field $\vec{E}_D$ that is
phase conjugated to $\vec{E}_{W^{\prime}}$. This field is
generated by the medium in the transient excitation of the
$\sigma_{1a,2}$ coherence, corresponding to the extraction of the
stored coherence grating. Using the stored state as initial
conditions, the solution of the above equations for
$\sigma_{1a,2}(t)$ is
\begin{align}
\sigma_{1a,2}(t) &=
\frac{\Omega_R^*\rho_{1a,1b}^s(t_s)e^{-\gamma_1 t}{\rm
senh}\left(\gamma_2\,t\right)}{\gamma_2}\;, \label{rho23t}
\end{align}
with
\begin{subequations}
\begin{align}
\gamma_1 &= \frac{\Gamma_{12} + \gamma}{2}, \\
\gamma_2 &=
\frac{\sqrt{\left(\Gamma_{12}-\gamma\right)^2-4|\Omega_R|^2}}{2},
\end{align}
\end{subequations}
The single-atom polarization vector $\vec{p}_{2,1a}$ on the $2
\rightarrow 1a$ transition is then given by
\begin{equation}
\vec{p}_{2,1a}(\vec{r},t) = \vec{d}_{2,1a}
\sigma_{2,1a}(\vec{r},t) e^{-i\omega_e t} \;.
\end{equation}

\subsection{Signal}

The electric field $\vec{E}_D$ for the $D$ field coming from the
diffraction of $\vec{E}_R$ on the sample coherence grating (see
Fig.~\ref{setup}b) is a result of the constructive interference of
the emission of all atoms in the $-\vec{k}_{W^{\prime}}$
direction. If we neglect interaction between atoms and propagation
effects on the $D$ field, again for simplicity, the value of
$\vec{E}_D$ in the $\vec{k}$ direction can be obtained by the
superposition of all atomic contributions on that direction:
\begin{equation}
\vec{E}_D(\vec{k},t) = \frac{1}{4\pi\epsilon_0 (2\pi)^{3/2}} \int
\eta(\vec{r}) \vec{p}_{2,1a}(\vec{r},t) e^{-i\vec{k}\cdot\vec{r}}
d^3\vec{r} \,, \label{ED_01}
\end{equation}
where $\eta (\vec{r})$ represents the atomic density at $\vec{r}$,
$\epsilon_0$ is the vacuum permittivity, and the integration runs
over the whole ensemble volume. Approximating the fields $W$,
$W^{\prime}$, and $R$ as plane waves, we can neglect the spatial
dependence on ${\cal E}_W$, ${\cal E}_{W^{\prime}}$, and ${\cal
E}_R$, respectively. In this case, we can write
\begin{subequations}
\begin{align}
\frac{\Omega_W}{\Gamma_{12}} &= i \sqrt{\frac{I_W}{2 I_{sa}}} e^{ik_W z} \;, \\
\frac{\Omega_{W^{\prime}}}{\Gamma_{12}} &= i \sqrt{\frac{I_{W^{\prime}}}{2 I_{sb}}} e^{i\vec{k}_{W^{\prime}}\cdot\vec{r}} \;, \\
\frac{\Omega_R}{\Gamma_{12}} &= i \sqrt{\frac{I_R}{2 I_{sb}}}
e^{-ik_R z} \;,
\end{align}
\end{subequations}
with $I_W$, $I_{W^{\prime}}$, and $I_R$ the intensities of the
$W$, $W^{\prime}$, and $R$ fields, respectively. $I_{sa}$ and
$I_{sb}$ are the saturation intensities of the $1a \rightarrow 2$
and $1b \rightarrow 2$ transitions, respectively, defined
according to Ref.~\onlinecite{Steck}.

Since $k_R - k_W = 0$, Eq.~\eqref{ED_01} can be written as
\begin{equation}
\vec{E}_D(\vec{k},t) = \frac{i \vec{d}_{2,1a}|\rho_{1a,1b}^s|
f_R(t)e^{-i\omega_e t}}{4\pi \epsilon_0 (2\pi)^{3/2}} \int
\eta(\vec{r}) e^{-i(\vec{k}+\vec{k}_{W^{\prime}})\cdot\vec{r}}
d^3\vec{r} \,, \label{ED_02}
\end{equation}
with
\begin{equation}
|\rho_{1a,1b}^s| = \frac{\sqrt{I_W I_{W^{\prime}}}e^{-\gamma
t_s}}{I_W\displaystyle\sqrt{\frac{I_{sb}}{I_{sa}}} +
I_{W^{\prime}}\sqrt{\frac{I_{sa}}{I_{sb}}}}\;
\end{equation}
representing the modulus of the stored ground-state coherence, and
\begin{equation}
f_R(t) = \sqrt{\frac{I_R}{2 I_{sb}}}\frac{e^{-\gamma_1\, t}{\rm
senh}\left(\gamma_2\, t\right)}{\,\gamma_2/\Gamma_{12}}
\end{equation}
a function describing the temporal profile of the D-field pulse.
Note that $f_R(t)$ is a function of the read field parameters
only.

If we approximate the distribution of atoms as having a gaussian
profile with the same rms width $L$ in all three direction, we can
write
\begin{equation}
\eta(\vec{r}) = \frac{N}{(2\pi L^2)^{3/2}}
e^{-\vec{r}\cdot\vec{r}/2L^2}\,,
\end{equation}
where $N$ is the total number of atoms in the cloud. Using this
expression for $\eta (\vec{r})$, Eq.~\eqref{ED_02} becomes
\begin{equation}
\vec{E}_D(\vec{k},t) = \frac{i \vec{d}_{2,1a} N |\rho_{1a,1b}^s|
f_R(t)e^{-i\omega_e t}}{4\pi\epsilon_0 (2\pi)^{3/2}}
e^{-|\vec{k}+\vec{k}_{W^{\prime}}|^2L^2/2} \,, \label{ED_03}
\end{equation}
which explicitly shows that the emission of the $D$-field occurs
in the $-\vec{k}_{W^{\prime}}$ direction only with a spread in
vector space, on each direction, of the order of the inverse of
the atomic-distribution spatial width, $L^{-1}$.

The detection apparatus can be arranged to collect all light in
the $D$-field mode. In this case, and if the detection of the
field is performed with a fast detector compared to the time
variation of $f_R(t)$, the signal $S_{\rm fast}(t)$ is then
proportional to the integration of the intensity of light in field
$D$ over all $\vec{k}$:
\begin{equation}
S_{\rm fast}(t) = A \int |\vec{E}_D(\vec{k},t)|^2 d^3 \vec{k} \;,
\end{equation}
where $A$ is a proportionality constant. From Eq.~\eqref{ED_03},
we see that such detected signal is given by
\begin{equation}
S_{\rm fast}(t) = A^{\prime} |\rho_{1a,1b}^s|^2 |f_R(t)|^2 \;,
\end{equation}
with $A^{\prime}$ a different proportionality constant.

Another important quantity that can be directly derived from
$S_{{\rm fast}}(t)$ is the total energy, $U_D$, extracted in mode
$D$. Note that, in light-storage measurements, the goal is usually
to extract as much information and energy as possible from the
coherence grating~\cite{Laurat_2006}. From the expressions derived
above we have then
\begin{align}
U_D &= \int_0^{\infty} S_{{\rm fast}}(t) dt \nonumber \\
&= \frac{2\,A^{\prime} |\rho_{1a,1b}^s|^2}{\Gamma_{12}} \frac{I_R
/ 2I_{sb}}{(1 + \frac{\gamma}{\Gamma_{12}})(\frac{I_R}{2I_{sb}} +
\frac{ \gamma}{\Gamma_{12}})} \;.
\end{align}

\section{Experimental results and discussions}

As indicated in Fig.1a the experiment was performed using a
degenerate two-level system. This system corresponds in the
experiment to the cycling transition $6S_{1/2}(F=3)\leftrightarrow
6P_{3/2}(F^{\prime}=2)$ of the cesium $D_{2}$ line. The cesium
atoms were previously cooled in a MOT operating in the closed
transition $6S_{1/2}(F=4)\leftrightarrow 6P_{3/2}(F^{\prime }=5)$
with a repumping beam resonant with the open transition
$6S_{1/2}(F=3)\leftrightarrow 6P_{3/2}(F^{\prime }=3)$. To prepare
the atoms in the state $6S_{1/2}(F=3)$, we switch off the
repumping beam for a period of about 1 ms to allow optical pumping
by the trapping beams via non resonant excitation to the excited
state $F^{\prime }= 4$. After optical pumping, the optical density
of the sample of cold atoms in the $F=3$ ground state is
approximately equal to 3 for appropriate MOT parameters.

All the incident laser beams indicated in Fig.1 are provided by an
external cavity diode laser which is locked to the
$F=3\leftrightarrow F^{\prime}=2$ transition. The grating writing
beams ($W$ and $W^{\prime}$) have the same frequency. After
passing through a pair of acousto-optical modulators (AOM) with
one of them operating in double passage, they can have their
frequency scanned around the $F=3\leftrightarrow F^{\prime}=2$
transition. The two AOM's also allow us to control their
intensity. These two beams are circularly polarized with opposite
handedness and are incident in the MOT forming a small angle
$\theta \approx ~60$ mrad, which leads to a polarization grating
with a spatial period given by $\Lambda=\frac{\lambda}{2sin(\theta
/2)}$, where $\lambda$ is the light wavelength. The reading beam R
is circularly polarized opposite to the writing beam W and also
passes through another pair of AOM's which does not change its
frequency but allow us to control its intensity.

 Employing the time sequence shown in Fig. 1c we have investigated
the light grating storage dynamics through the observation of
delayed Bragg diffraction of the reading beam R in the Zeeman
coherence grating induced by the writing beams $W$ and
$W^{\prime}$. The writing and reading pulses are trigged to the
switching off of the repumping laser which also triggers the turn
off of the MOT quadrupole magnetic field. In order to compensate
for spurious magnetic fields, three independent pairs of Helmholtz
coils with adjustable currents are placed around the MOT.

\begin{figure}
\includegraphics[angle=-90,scale=0.35]{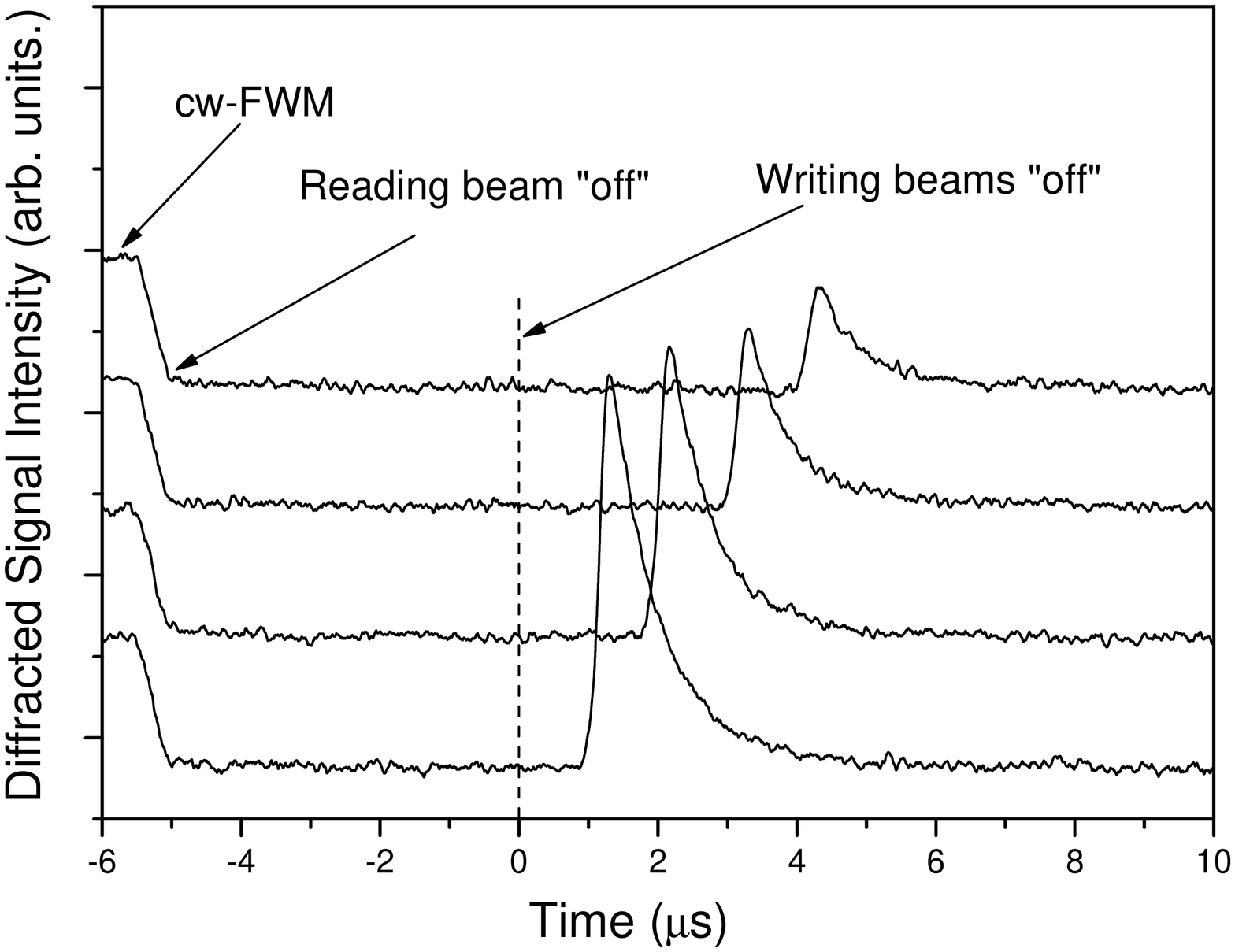}
\caption{Bragg diffraction signal retrieved from the stored
grating for different storage times}
\end{figure}

\begin{figure}
\includegraphics[angle=-90,scale=0.35]{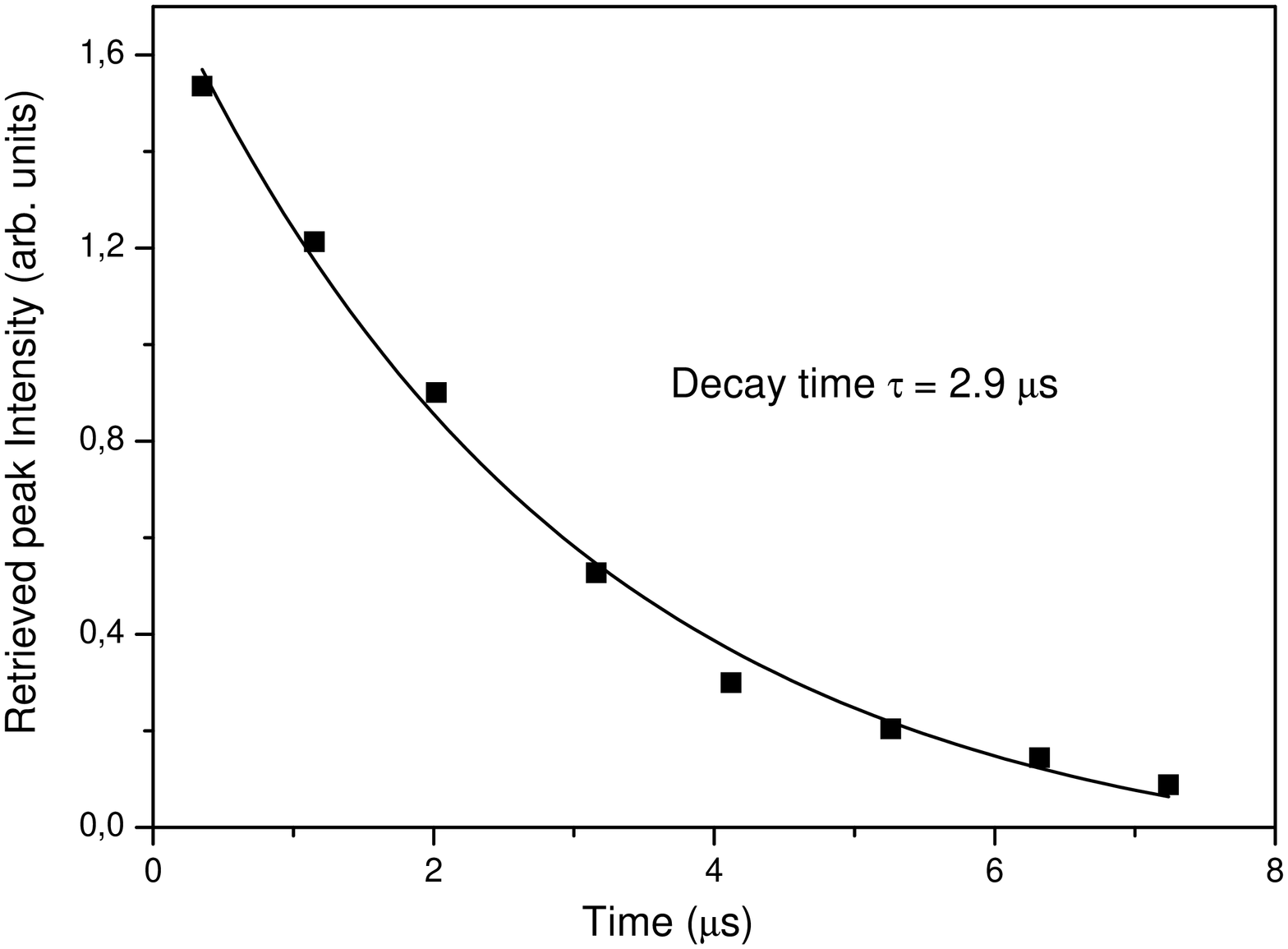}
\caption{Normalized Bragg diffraction peak signal for different
storage times. The solid curve corresponds to a fit with an
exponential function.}
\end{figure}

\begin{figure}
\includegraphics[angle=-90,scale=0.32]{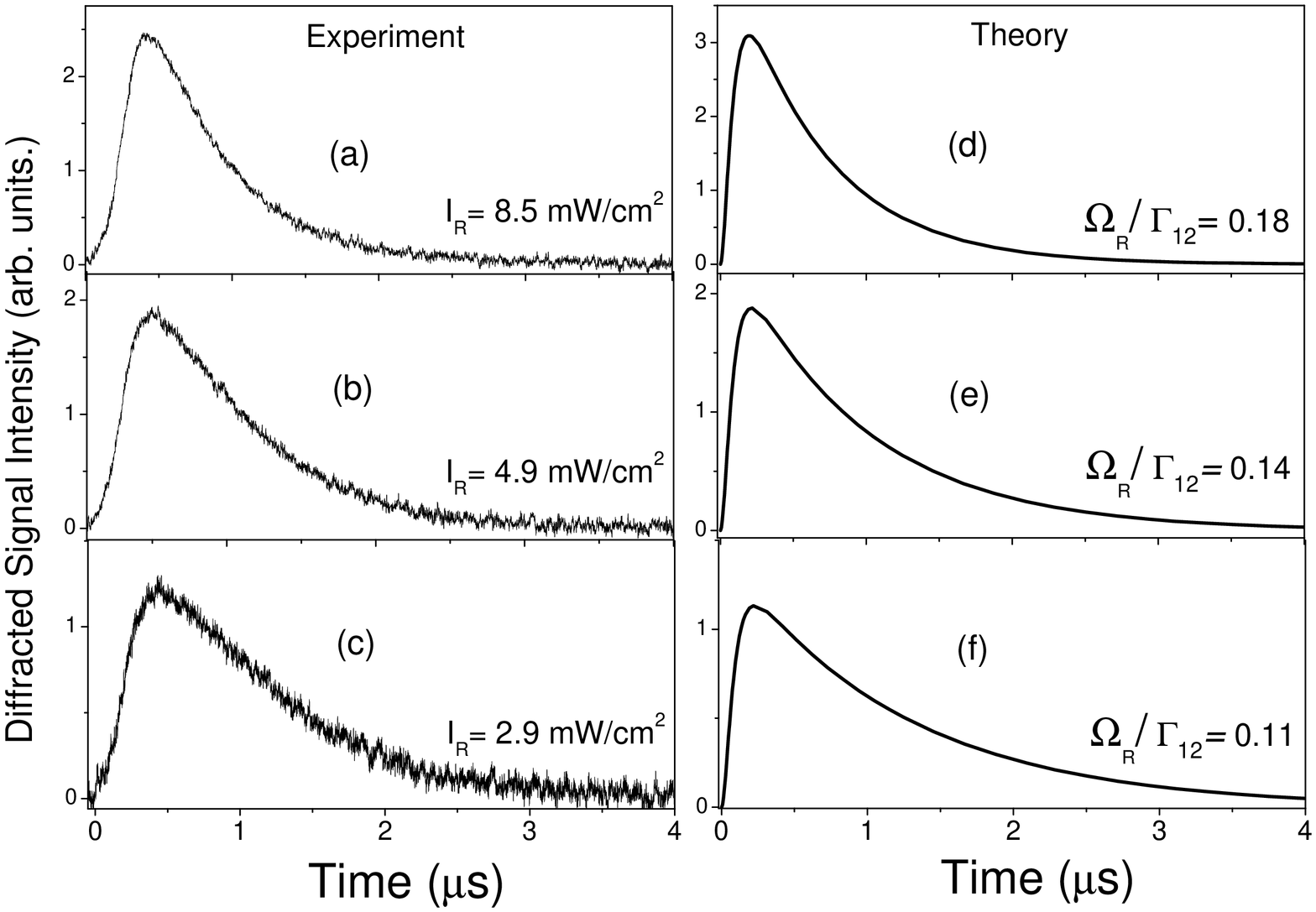}
\caption{Experimental (right) and calculated (left)
retrieved-pulse temporal shape corresponding to different
intensities of the reading beam, for a fixed storage time. The
writing beam intensities, $W$ and $W^{\prime}$,  are 7.0 mW/cm$^2$
and 1.0 ~mW/cm$^2$ respectively. As described in the text, for
comparison between theory and experiment, all the theoretical
reading beam intensities needs to be re-scaled by a factor
$a\approx 0.02$.}
\end{figure}

In Fig. 2 we show the cw-FWM and the Bragg diffracted signal which
is retrieved from the stored Zeeman coherence grating for
different storage times. We have experimentally verified that the
polarization of the diffracted beam, both for the steady state
cw-FWM signal (real time Bragg diffraction) and for the retrieved
signal, is always opposite to the polarization of the reading beam
as schematically depicted in Fig. 1b. We have been able to observe
the diffracted signal up to a time of $10 ~\mu $s. This maximum
storage time is very sensitive to the compensation of the residual
magnetic field. It is interesting to note that for short storage
times the retrieved signal peak intensity is much larger than the
corresponding cw-FWM signal. This effect is related to the
simultaneous presence of the writing and reading beams in the cw
regime, where the reading beam contributes to decrease the
contrast of the coherence grating induced by the writing beams.
The decay of the peak intensity of the diffracted pulse,
normalized by its steady state value (cw-FWM signal) is presented
in Fig. 3. The exponential decay behavior is evidenced by the
exponential fitting (solid curve). For the data presented in Fig.
3, the intensities of the writing beams $W$ and $W^{\prime}$ are
approximately equal to 5.0 mW/cm$^2$ and 1.5 mW/cm$^2$
respectively, while the intensity of the reading beam R is about
8.0 mW/cm$^2$. From the measurement presented in Fig. 3, we obtain
a decay time of the order of $2.9 ~\mu$s, which corresponds to the
Zeeman ground state coherence decay. It is worth noticing that we
have experimentally verified that the measured coherence time does
not depend on the intensity of either the writing and the reading
beams.

For a fixed storage time of approximately ~$1 \mu$s, we have also
measured the temporal pulse shape of the retrieved signal for
different reading beam intensities and the results are shown in
Fig. 4a-c for three different values of the reading beam
intensity. We note that the experimentally retrieved pulse raising
time is limited by the time constant of the detector ($\leq
0.5~\mu$s). As we have discussed previously the coherently
prepared atomic system couples to the reading beam to transiently
generate the diffracted pulse signal. The temporal width of the
generated pulse decreases for increasing reading beam intensity, a
direct consequence of the effect of the increased dumping of the
Zeeman ground state coherence caused by spontaneous emission
induced by the reading beam in the process of mapping the stored
Zeeman coherence into the optical coherence. In Fig. 4d-f we show
the corresponding retrieved pulse obtained using the previously
developed theory, assuming $I_{sb}$ the saturation intensity for
the $6S_{1/2}(F = 3 ,m_F = +3) \rightarrow 6P_{3/2}(F = 2 ,m_F =
+2)$ transition. We have used an adjustable parameter of the order
of $a\approx 0.02$ to re-scale all the theoretical reading beam
intensities (i. e., $I_{R}\rightarrow aI_{R}$), which accounts for
the uncertainty in the determination of the effective experimental
value of the Rabi frequency associated with the reading beam.

More systematically, in Fig. 5 we plot the measured pulse width
(FWHM) for different reading beam intensities. In these
measurements, for each value of the reading beam intensity, we
have recorded three curves of the retrieved pulse which allows us
to estimate the corresponding error bars. The solid curve in Fig.
~5 corresponds to the calculation of the pulse temporal width
using the signal shape function given by $Eq.~25$. In this
calculation we have used $\gamma/\Gamma_{12}\approx 0.014$ in
order to obtain the best agreement with the experiment. Note that
this value is of the same order of the experimentally measured
decay rate, obtained from the different set of data shown in Fig.
3, and estimated as $\gamma/\Gamma_{12}\approx 0.02$, with
$2\Gamma_{12}/2\pi=\Gamma_{22}/2\pi=$ 5.2 MHz. From the same set
of data as Fig.~5, we show in Fig. 6 the retrieved pulse energy,
obtained by time integration of the measured pulse intensity. The
corresponding solid curve is a theoretical fitting obtained using
$Eq.~26$ with the same adjustable parameter $a$. As can be
observed the agreement between theory and experiment is
qualitatively satisfactory owning to the simplification of the
theoretical model, which consider a single three-level system and
do not accounts for the manyfold Zeeman degeneracy. In fact, for
the intensities of the writing beams used in the experiment,
ground state coherence involving different pairs of Zeeman
sub-levels can actually exist.

\begin{figure}
\includegraphics[angle=-90,scale=0.3]{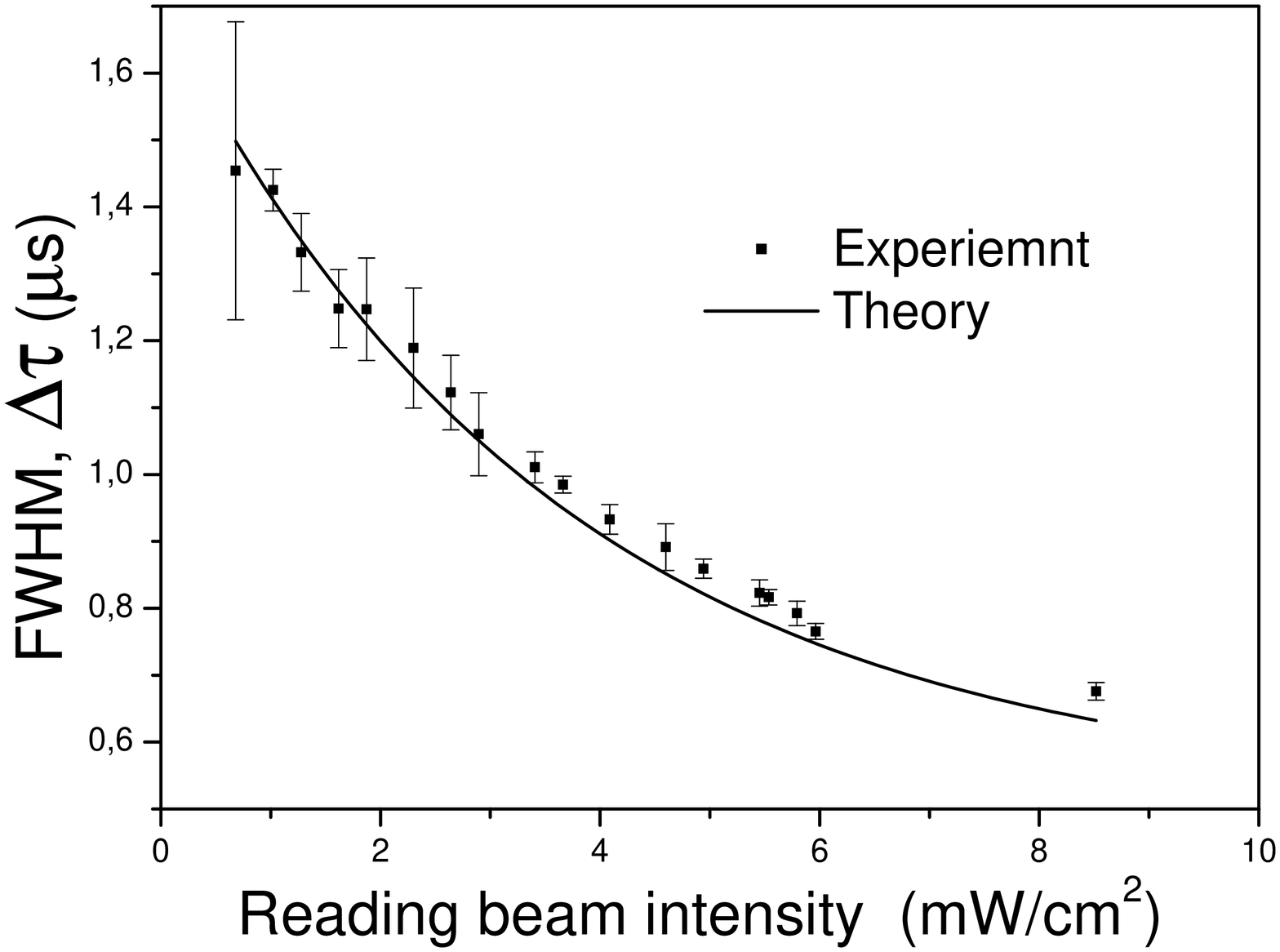}
\caption{Measurement of the temporal width (Full Width at Half
Maximum) of the retrieved pulse for different intensities of the
reading beam, obtained in similar experimental conditions as in
Fig. 4a-c. The solid curve is a theoretical fitting using the
model described in the text.}
\end{figure}

\begin{figure}
\includegraphics[angle=-90,scale=0.3]{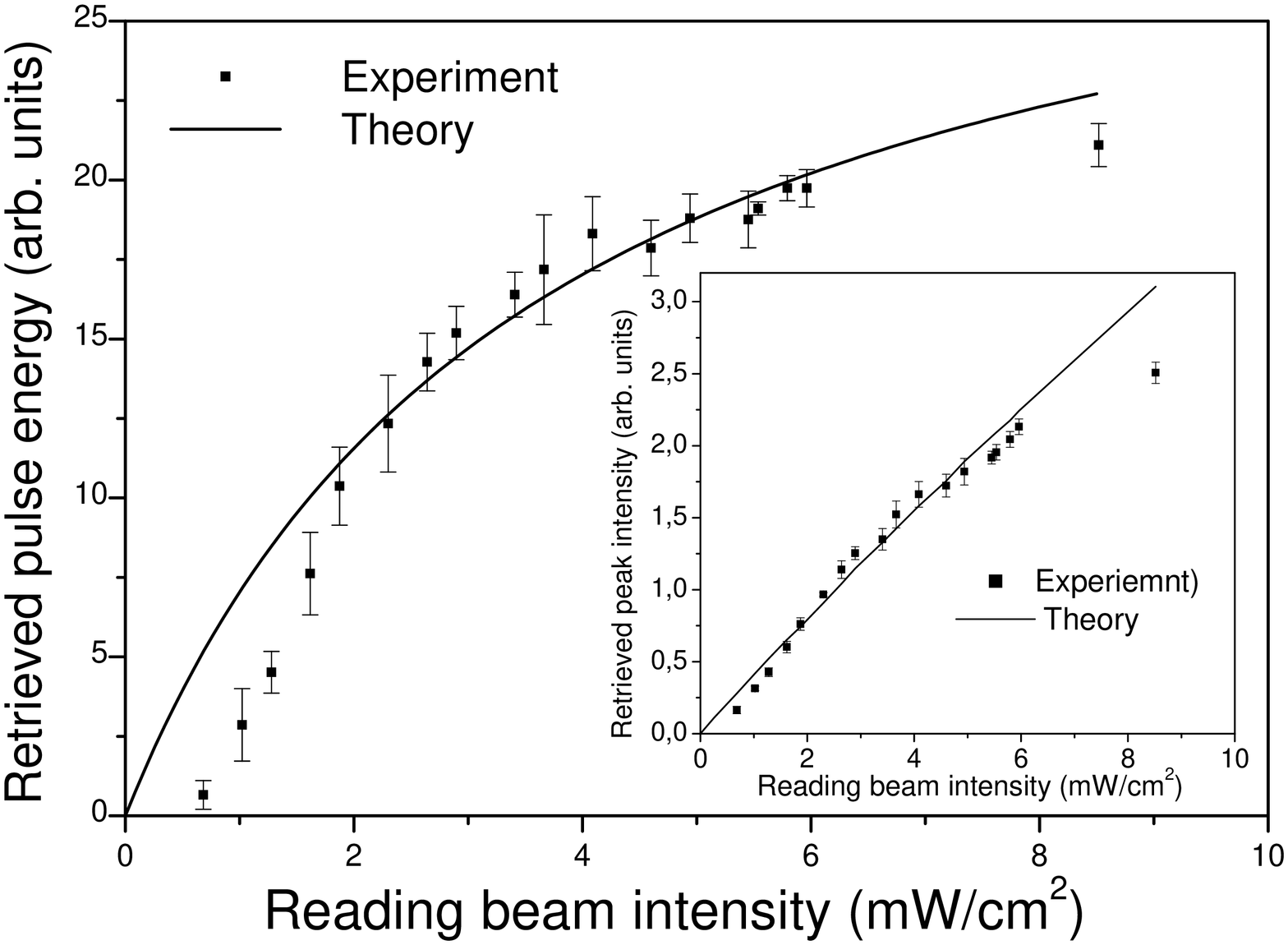}
\caption{Retrieved pulse energy for different intensities of the
reading beam, obtained on similar experimental conditions as in
Fig. 4(a-c). Inset: The corresponding variation of the peak
intensity of the retrieved pulse. The solid curves are theoretical
fittings employing the model described in the text with the same
intensity adjust parameter used to fit the pulse width in Fig.~5.}
\end{figure}

\begin{figure}
\includegraphics[angle=-90,scale=0.3]{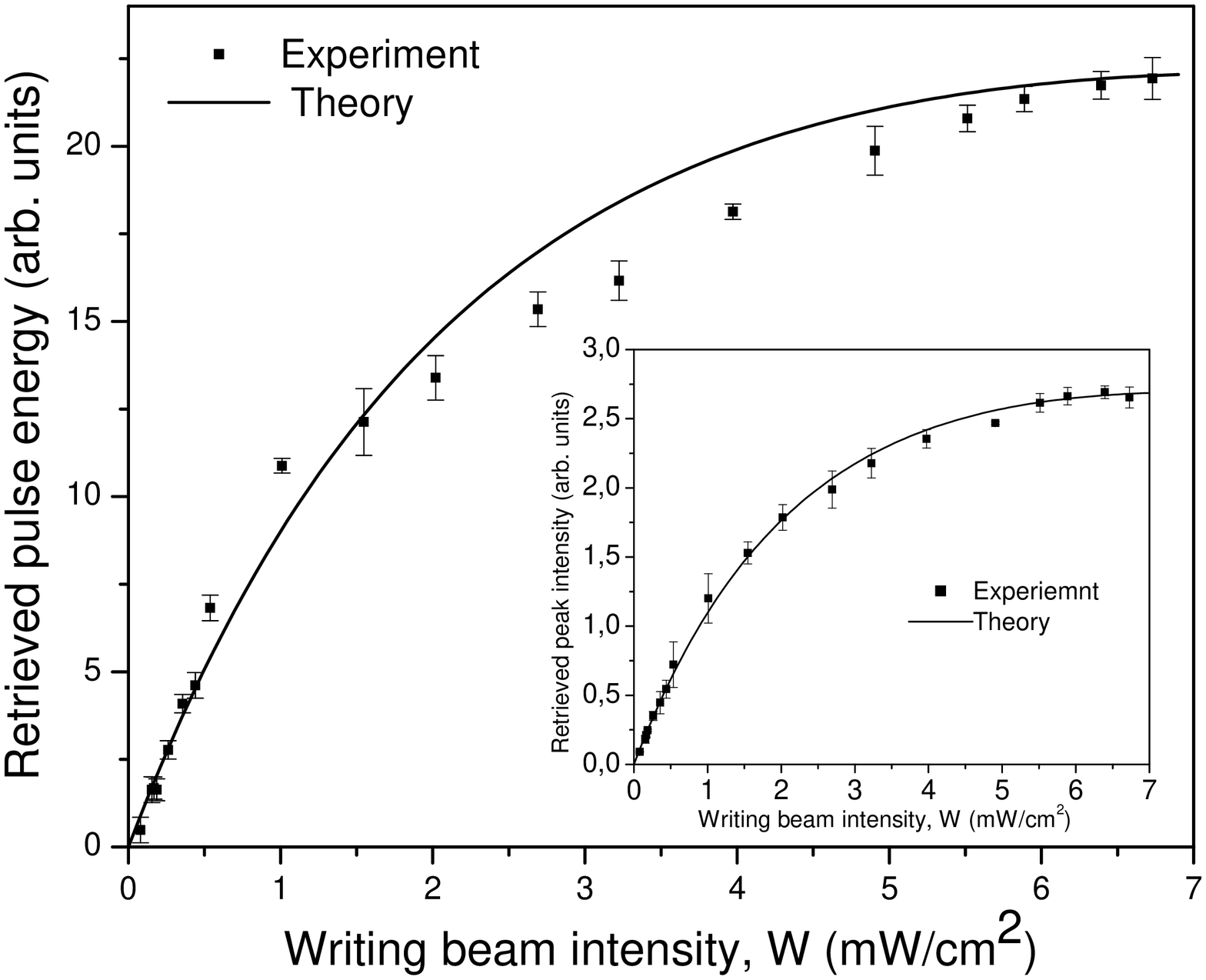}
\caption{Variation of the maximum peak intensity of the retrieved
pulse for different intensities of the grating writing beam W. For
these data, the corresponding intensities of the writing
($W^{\prime}$) and reading (R) beams were fixed at 1.0 mW/cm$^2$
and 9.0 mW/cm$^2$, respectively. The solid curves are again
theoretical fittings using the model described in the text. We
have used the same intensity adjust parameter, $a^{\prime}\approx
1.9$, in both curves.}
\end{figure}

We also have measured the variation of the diffracted signal as a
function of the intensity of one of the grating writing beam
(i.e., the beam $W$) and the results for the corresponding pulse
energy are shown in Fig. 7. For these measurements, the
intensities of the grating writing beam $W^{\prime}$ and the
reading beam were respectively equals to 1.0 mW/cm$^2$ and 9.0
mW/cm$^2$. The solid curve in Fig. 7 corresponds to a theoretical
fitting with the calculated retrieved pulse energy given by
$Eq.~26$, assuming that $I_{sa}$ is the saturation intensity of
the $6S_{1/2}(F = 3 ,m_F = +1) \rightarrow 6P_{3/2}(F = 2 ,m_F =
+2)$ transition, with $I_{sa}=15I_{sb}$ according to the ratio
between the corresponding Clebsch-Gordan coefficient. Again, to
account for the uncertainty in the experimental value of the Rabi
frequency associated with the writing beams $W$ and $W^{\prime}$,
we have used another adjustable parameter, which in the present
case is of the order of $a^{\prime}\approx 1.9$, to re-scale the
theoretical intensity ratio between these beams (i.e.,
$I_{W}/I_{W^{\prime}}\rightarrow a^{\prime}
I_{W}/I_{W^{\prime}})$.

As can be observed from Fig. 6 and Fig. 7 the amount of energy
that can be retrieved from the medium clearly saturates with the
writing and reading beam intensities. In particular, this shows
that for fixed writing beams intensities, there is a maximum
amount of energy that can be retrieved from the stored coherence.
However, it is worth mentioning the different saturation behavior
observed for the variation of the peak intensity of the retrieved
pulse with the corresponding intensity, as shown in the insets of
Fig.~6 and Fig.~7. As observed, the maximum peak of the retrieved
pulse saturates more strongly with the writing beam intensity as
compared with the saturation induced by the reading beam. As one
should expect, the saturation induced by the reading beam is
related mainly to the total retrieved energy. The pulse peak,
however, can increase much further with the reading power, since
it is closely related also with the speed of the reading process.
On the other hand, the increase of the writing beam intensity will
saturate the Zeeman coherence grating, therefore reducing its
contrast. This effect has a strong influence on the Bragg
diffraction efficiency, affecting equally the total retrieved
energy and the pulse peak.

\section{Summary}

We have investigated, both theoretical and experimentally, the
storage of a spatial light polarization grating into the Zeeman
ground states coherence of cold cesium atoms. Systematic
measurements were performed to reveal the saturation behavior of
the retrieved signal as a function of the intensities of the
writing and reading beams. The developed simple theoretical model
accounts reasonably well for the observed results and in
particular for the measured pulse temporal shape. We consider our
results are of considerable importance for a better understanding
of the coherent memory for multidimensional state spaces. Finally,
we would like to mention that we also have observed the coherent
evolution of the stored grating in the presence of an applied
magnetic field, which shows collapses and the revivals of the
stored coherence grating. This effect is associated with the
Larmor precession of the induced grating around the applied
magnetic field as was reported previously in
\cite{Matsukevich06,Jenkins06} and strongly support the
possibility of manipulating more complex spatial information
stored into an atomic medium. Further investigation on this effect
is currently under way and will be published elsewhere.

\vspace{0.3 cm}

We gratefully acknowledge Marcos Aurelio for his technical
assistance during the experiment. This work was supported by the
Brazilian Agencies CNPq/PRONEX, CNPq/Inst. Mil\^{e}nio and FINEP.

\end{document}